\documentclass[prd,aps,preprint,tightenlines,superscriptaddress]{revtex4}
\usepackage{mathrsfs}
\usepackage{amssymb}
\usepackage{color}
\usepackage{epsfig}
\usepackage{graphicx}
\usepackage{amsmath}
\begin{document}
\title{Dark Matter Tomography}
\author{Shmuel Nussinov}
\affiliation{Tel Aviv University, Israel and Chapman College, California}


\begin{abstract}
We consider Wimp annihilations into monochromatic and continuous $\gamma$'s and the angular  distribution of the resulting gammas. We discuss how the WIMP density profile can be reconstructed from the angular dependence
of the photon flux.
\end{abstract}
\maketitle

\section{Introduction}
Finding direct evidence for cold dark matter (CDM) which may comprise $25\%$ of the
cosmological energy density and the missing matter in galactic halos became a ``holy
grail" for many experimental searches.

There is also the exciting prospect that the weakly interacting massive particles(WIMPs)
denoted here by $\chi$ and $\bar \chi$  comprising the CDM hail the new TeV
scale physics responsible for the electroweak symmetry breaking.

In Supersymmetry (SUSY), Technicolr and other scenarios beyond the standard model,  the WIMP masses are often in the
100 GeV - TeV mass scale and may manifest via energetic photons emerging from $\chi - \bar\chi$
annihilations in overdense regions in our halo.

In some models the WIMPs are unstable and their decays produce energetic $\gamma$'s.
Even long lifetimes consistent with the role of WIMPs as dark matter can then yield
a non-negligible $\gamma$ signal.

In the following we make some general and model dependent statements on the detectability
of the these signals via their energy spectra and/or their directionality.
We also address the question of using, at some future time, precise measurements of these
signals to perform a``Tomography" reconstructing of the spatial WIMP density distribution.

\section{Some General Comments}

In certain models (see e.g Ref. by us for decay and by Ulio Bergstrom, Edso
and Gondolo for annihilation..) the WIMP decay or annihilate with significant
branchings into a photon and another stable particle or narrow resonance:
\begin{equation}
\chi (\; \rm{or}\;\bar\chi-\chi) \to \gamma  +  h^0.
\end{equation}
In such cases the photon is almost monochromatic.

More generically the annihilation/decay yields several particles including one (or
more) photons. Specific examples are the annihilation:
\begin{equation}
\chi + \bar {\chi} \to U + U,\; U \to \pi^+ \pi^- \pi^0,\; \pi^0 \to 2\gamma;
 \end{equation}
or the decay:
\begin{equation}
 \chi \to \nu \tau^+ \tau^-,\; \tau^- \to \nu \pi ^- \pi^0,\; \pi^0 \to 2\gamma;
\end{equation}
The $\gamma$ signal in such cases is broad:
\begin{equation}
\Delta (E_{\gamma}) \sim \langle E_{\gamma}\rangle \sim f m_\chi,
\end{equation}
where $f$ (=1/{12} or 1/{18} in the above examples) is the average fraction of the initial
mass in a  photon. Clearly it is a less striking signal than monochromatic $\gamma$'s.
Yet the quick rise towards a multi-Gev peak and subsequent fall to zero, expected here,
are distinct from astrophysical signals which usually are monotonically decreasing,
often with power like fall-off, and no sharp cut-off.

To date we have no evidence for WIMP annihilation /decay yielding $\gamma$ signals.
Indeed certain models with (Sommerfeld enhanced) WIMP annihilations accounting for possible
deviations from standard astrophysical signals of electros /positrons are constrained by
requiring that annihilations as in {Eq.2} above will not yield a corresponding enhanced
photon flux from the galactic center~\cite{Meade:2009rb}.
Discovering in future, more sensitive experiments, a monochromatic $\gamma$ line in the
100 GeV -TeV mass range will practically establish WIMPs.

In the following we will assume that a monochromatic line or a multi-GeV peak were found
in the $\gamma$ spectrum and tentatively associated with WIMPs.

We generally expect that the directional distribution of such $\gamma$'s will peak to
varying degrees in the direction of the galactic center. Conversely precise future
measurements of the angular dependence of the WIMP associated $\gamma$ flux relative
to the galactic center and the galactic plane provide a Tomographic images of the
WIMP's density distribution. Such mesurements allow finding $\rho (\vec {r'})$, the
density distribution of WIMP's .

The cases of $\gamma$'s originating from annihilations/decays are rather different
both phenomenologically and from the underlying physics points of view.

Specifically, as we list below:

(i). Both annihilations and decays can occur in scenarios where WIMP's are freeze
-out remnants of cosmological early annihilations (at times when the temperature T
is about $(1/{20}\sim 1/{30}) m_\chi$) .

It is however conceivable that dark matter arises, just like ordinary baryons, after
efficient annihilations left only the excess of ``matter" (say). By definition in such
``asymmetric" scenarios we have no WIMP annihilation at present. Still if the ``charge"
associated with this matter is, just like baryonic charge is not exactly conserved
WIMP decays are possible.

Baryonic decays are very slow ($\tau _{B} > (10^{31}$ years). However the
situation can be different in ``asymmetric" WIMP models . Thus in a Technibarionic analog
~\cite{Nussinov:1985xr}  with technibaryon mass $m_{TB} \sim 10^3 m_{B}$ the decay rate, 
scaling as $m^5$, may be ${\mathcal {O}} (10^{16}$ years)$^{-1}$, generating an 
observeable $\gamma$ signal!

(ii). WIMP decays can occure also in symmetric WIMP models such as those with slowly decaying
gravitinos~\cite{gravitino}. Finally in all cases sufficiently strong WIMP
nuclear interactions allow direct observation via nuclear recoil in underground detectors.

(iii) For WIMPs decaying (but {\it  not} annihilating) into $\gamma$'s there is an amusing
`` Cosmological Echo" of the ``Local" signal from WIMPs decaying in the halo.
The intensity of the $\gamma$ fluxes are given by the line integrals of the density of WIMPs:
\begin{equation}
 \Phi_{local}\sim \rho_{local}\times R_{haloe} \sim 0.3\rm {GeV/{(cm)^3}}\times  30\rm{kpc}
\end{equation}
 and
\begin{equation}
\Phi_{cosmological} \sim \rho_{cosmo} \times R_{Hubble} \sim 3 \rm{keV/{(cm)}^3}\times 3 \rm{Gpc}
\end{equation}
respectively. The spurious similarity of the two fluxes suggests that both are jointly (un)observeable.

(The Hubble expansion dilutes the cosmological flux from high red-shifts by $1/{(1+z)^3}$.
Also the WIMP mass density is $\rho \sim 0.3$ GeV/{cm$^3$}  at our neighbourhood, falling at
distances $|r'| \gg a \sim 8$ kpc  from the galactic center at least as  $\sim~1/{|r'|^2}$.)
Yet there are quite significant differences in energy spectrum and directional distribution
which may allow resolving the two signals.

The local signal originating from decays of halo WIMPs {\it  can} be  monochromatic as the
Doppler broadening due to the virial WIMP velocities therein is of order $0.1 \%$ only.
This is not so for $\gamma$'s from decays of cosmological WIMPs which are redshifted by a 1+z
factor. Also the cosmological signal is isotropic whereas due to the higher density of WIMPs
towards the galactic center, the flux of $\gamma$'s from their decay of is unisotropic, enhanced
in the direction of the galactic center. For  WIMPs decaying into $n > 2$ bodies with only a peak
in the spectrum the effect of the cosmological redshift is more difficult to ascertain .

Still a mild softening of the $\gamma$ spectrum from directions further away from the galactic
center may indicate the blending in of the cosmological component.

(iv)  The signal due to WIMP annihilation from any given direction is proportional to the
line integral of the ${\it square}$ of the number density of WIMPs along that direction.
Thus unlike $\gamma$'s from decays to which cosmological and haloe WIMPs make similar contri-
butions, annihilating of haloe WIMPs dominate cosmological WIMPs by the ratio of WIMP densities:
\begin{equation}
 (0.3 \rm{GeV/{cm^3}})/ (3 \rm{keV/{cm^3}})\sim 10^5\,.
\end{equation}

Further the much stronger variation of the volume emissivity as a function of $|r'|$ , the
distance from the galactic center generates a more pronounced unisotropy and enhanced flux
from the galactic center than in the previouse case of decaying WIMPs.

 (v).  For the exothermic annihilation process $ v\cdot\sigma_{annihi}$ remains generally
constant in the threshold region even as the relative velocity $v$ tends to zero. Thus the rate
of annihilations per unit time and volume (and resulting volume emmisivity of photons)
which are proportional to $n^2\cdot v\cdot\sigma_{annihi}$ is indeed proprtional to the square
of the WIMP number density as stated above.

However in models with extra, light vectors (or scalars) (termed $U$ bosons) the annihilation of WIMPs is further
accelerated by the ``Sommerfeld enhancement"  (see~\cite{ArkaniHamed:2008qn}).
It introduces an extra factor of $\pi\cdot\alpha'/v$ with $\alpha'$ the analog of the electromagnetic
$\alpha \sim 1/{137}$ and for virial speeds $\sim 10^{-3}$ and $\alpha' > \alpha$ can be quite substantial.
(The finite mass $m_{U}$ of the exchanged bossons provides an infrared cut-off saturating the
cross-sections at $\sim 2\pi/{m^2}$).

The Sommerfeld enhancement provides yet another preference for annihilations close to
the galactic center where the slowest WIMPs tend to ``sink" to. Simple energtics imply that:
\begin{equation}\label{velocity}
 \langle v(r')^2 \rangle = \frac{4\pi G_N}{r'} \int_{0}^{r'} \rho(r'') r''^2 dr''\,.
\end{equation}
Hence the velocity tends to decrease towards the galactic center and the Sommerfeld enhancement
increases therein.

\section{Reconstruction of Density Profiles from Measurements of the gamma Fluxes}

In all the three cases above there is a certain ``profile" $p(r')$ of the annihilation rate
and the ensuing $\gamma$ volume emissivity. Specifically we have
\begin{equation}
 p(r')\sim\rho(r')    ... \rm{case}\; A\,,
\end{equation}
 for decaying WIMPs
\begin{equation}
 p(r')\sim\rho(r')^2  .... \rm{case}\; B\,,
\end{equation}
 for annihilating Wimps, and
\begin{equation}
 p(r') \sim\rho(r')^2/{v(r')}  .... \rm{case}\; C\,,
\end{equation}
 with $v(r')$ the the avarage velocity defined by Eq.~(\ref{velocity}) above,  if we
have also Somerfeld enhancement.

It has been recently argued that while the overall WIMP
density $\rho(r')$ with $r'$ the distance from the galactic center is spherically symmetric,
the velocity distribution is not. Rather they suggest that the velocity distribution (and
ensuing source profile $p(\vec{r'})$ in case C above) are modified by the disc and can no longer
be spherically symmetric.

The task tackle here is the reconstructing $p(\vec{ r'})$ the from the
angular distribution of the $\gamma$ flux:
\begin{equation}
\Phi(\Omega) = \frac{dN(\gamma)}{d\cos(\theta)d(\phi)}
\end{equation}
where $r$, $\theta$ and $\phi$ defined in polar coordinates, the vector $\vec {r}$ from us to the source
point of interest and the $z$ and $x$ axes used in defining $\theta$ and $\phi$ are the direction to
the galactic center and the (rotation) axis of the disc respectively and $r'$ is the distance
from the galactic center.

In principle this can be done also when $p(\vec {r'})$ is not spherically symmetric thereby
inducing also an azimuthal $\phi$ angle dependence of the observed $\gamma$ flux.

 Consider first spherical profiles $p(r')$ and fluxes $\Phi(\theta)$ depending on $\theta$ only.
 To map $p(r')$ to $\Phi(\theta)$ we integrate $d(r)p(r')$ with $r'=(r^2+a^2-2 a r \cos\theta )^{1/2}$ 
 from 0 to infinity along the line of sight
 to us (the $r^2$ factor in the volume element $r^2 d\Omega$ cancels with the $1/r^2$ geometric
 divergence of the emitted $\gamma$'s):
\begin{equation}
 \Phi(z) = \int dr p(r')\;,
\end{equation}
 with $a \sim 8 $ kpc our seperation from the galactic center and $z=\cos\theta$.

\begin{figure}[hbt]
\begin{center}
\includegraphics[width=10cm]{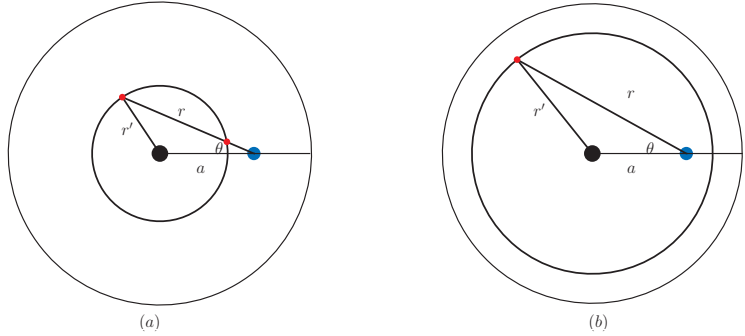}
        \caption{The dark halo contribution for the photon signals with azimuthal angle $\theta$.}
\end{center}\label{fig4}
\end{figure}

 Rather than trying to invert that relation to express $p(r')$ in termes of
$\Phi(z)$ we use fig. 1-a and fig. 1-b to illustrate the situations for $r'<a$ and for $r'>a$.
In the first case and  for $1-z^2 < (r'/a)^2$ the fixed r' circle is cut by the line of
sight at an angle $\theta$ to the G.C.at two points $r_{1,2}= az\pm (r'^2 -a^2 (1-z^2))^{1/2}$.

 In the second case there is always one such crossing at $r=az +[r'^2-a^2(1-z^2)]^{1/2}$
 Since the initial mapping of $p(r')$ to $\Phi(z)$ is linear in $dr$, the inverse mapping requires
 just the jacobian $|dr/{dr'}|$ , with $z$ held fixed at $r'/{[r'^2 -a^2(1-z^2)]^{1/2}}$, $= K(z,r')$
 hence, by integrating next over $dz$ we find that for $r' <a$ :
 \begin{equation}
 p(r') = \int_{0}^{\sqrt{1-(r'/a)^2}} { \Phi(z)2 K(z,r')}
\end{equation}
 and for $r'>a$ we have the same expression but with $1/2$ the magnitude.

\section*{Acknowledgments}
This paper was motivated as a sequel to a joint paper with S.L.Chen, R.N. Mohapatra and Y.Zhang in which the possibility of a monochromatic photon from a two body decay of an LSP gravitino occured. I am particularly indebted to S.L.Chen for his crucial help .

\end{document}